\documentclass[conference]{IEEEtran}
\IEEEoverridecommandlockouts
\usepackage{cite}
\usepackage{amsmath,amssymb,amsfonts}
\usepackage{algorithmic}
\usepackage{graphicx}
\usepackage{textcomp}
\usepackage{xcolor}
\usepackage{comment}
\def\BibTeX{{\rm B\kern-.05em{\sc i\kern-.025em b}\kern-.08em
    T\kern-.1667em\lower.7ex\hbox{E}\kern-.125emX}}
\usepackage[numbers]{natbib} 
\usepackage{capt-of}

\begin{document}

\title{GARCH-Informed Neural Networks for Volatility Prediction in Financial Markets}


\author{\IEEEauthorblockN{Zeda Xu}
\IEEEauthorblockA{\textit{College of Engineering} \\
\textit{Carnegie Mellon University}\\
Pittsburgh, PA, USA, 15213 \\
zedaxu@cmu.edu}
\and

\IEEEauthorblockN{John Liechty}
\IEEEauthorblockA{\textit{Smeal College of Business} \\
\textit{Pennsylvania State University}\\
University Park, PA, USA, 16802 \\
jcl12@psu.edu}
\and

\IEEEauthorblockN{Sebastian Benthall}
\IEEEauthorblockA{\textit{School of Law} \\
\textit{New York University}\\
New York City, NY, USA, 10012 \\
spb413@nyu.edu}
\and

\IEEEauthorblockN{Nicholas Skar-Gislinge}
\IEEEauthorblockA{
\textit{Lund University}\\
Lund, Sweden \\
nskg@pm.me}
\and

\IEEEauthorblockN{Christopher McComb}
\IEEEauthorblockA{\textit{College of Engineering} \\
\textit{Carnegie Mellon University}\\
Pittsburgh, PA, USA, 15213 \\
ccm@cmu.edu}
}

\maketitle

\begin{abstract}
    Volatility, which indicates the dispersion of returns, is a crucial measure of risk and is hence used extensively for pricing and discriminating between different financial investments. As a result, accurate volatility prediction receives extensive attention. The Generalized Autoregressive Conditional Heteroscedasticity (GARCH) model and its succeeding variants are well established models for stock volatility forecasting. More recently, deep learning models have gained popularity in volatility prediction as they demonstrated promising accuracy in certain time series prediction tasks. Inspired by Physics-Informed Neural Networks (PINN), we constructed a new, hybrid Deep Learning model that combines the strengths of GARCH with the flexibility of a Long Short-Term Memory (LSTM) Deep Neural Network (DNN), thus capturing and forecasting market volatility more accurately than either class of models are capable of on their own. We refer to this novel model as a GARCH-Informed Neural Network (GINN). When compared to other time series models, GINN showed superior out-of-sample prediction performance in terms of the Coefficient of Determination ($R^2$), Mean Squared Error (MSE), and Mean Absolute Error (MAE). 
\end{abstract}


\section{Introduction}
Prediction of market trends has long been an interest of the finance community \cite{henrique_literature_2019}. Numerous mathematical models have been proposed to fit and extrapolate historical data, with the goal of gaining exclusive insights into the future. Those models often assume that market occurrences will repeat themselves and, thus, the future is predictable based on historical data \cite{abu-mostafa_introduction_1996}. Stock price time series are noisy and volatile, and there is little to no predictability in the first moment of how a stock price changes over time \cite{abu-mostafa_introduction_1996}. There is, however, considerable structure and predictability in the second moment or volatility, which reflects the riskiness of a stock \cite{kim_forecasting_2018, markowitz_portfolio_1952}. Volatility prediction models aim to match and forecast the dispersion of the returns (typically on a log scale) of stock prices. These volatility predictions carry significant meaning and can inform investment decisions \cite{kim_forecasting_2018, koo_hybrid_2022}. 

One of the most widely used models for financial times series is the Autoregressive Conditional Heteroscedasticity (ARCH) model and many of its succeeding variations \cite{koo_hybrid_2022, engle_autoregressive_1982, alam_forecasting_2013}. It captures the heteroscedasticity (non-constant variance) that is present in most real-world data, enabling the estimation of current conditional variance based on previous values \cite{kim_forecasting_2018, engle_autoregressive_1982}. A generalized version of ARCH, GARCH (Generalized-ARCH) is a natural extension and widely considered as a standard model for forecasting stock volatility \cite{bollerslev_generalized_1986, hansen_forecast_2005, franses_forecasting_1996}. A large number of extensions of the GARCH models have subsequently been introduced, including the Exponential GARCH (EGARCH), the Glosten-Jagannathan-Runkle GARCH (GJR-GARCH) model, and the Threshold GARCH (TGARCH) model, all of which attempt to capture the leverage effect in financial time series with asymmetric volatility terms \cite{nelson_conditional_1991, glosten_relation_1993, zakoian_threshold_1994, kluppelberg_continuous-time_2004, li_zd-garch_2018}. Overall, the GARCH model and its extensions have shown decent performance in financial volatility forecasting tasks, but the performance of the models isn’t generalizable to all market conditions, and they are known to fail in certain market index forecasting applications \cite{franses_forecasting_1996, liu_volatility_2009, lee_are_1991, mcmillan_forecasting_2000}. This can be attributed to the fact that models in the ARCH family are highly linear, and therefore lack the capability to capture non-linear market features \cite{koo_hybrid_2022, lee_are_1991}. This characteristic hinders their prediction performance in out-of-sample time series prediction tasks. 

Machine Learning (ML) models have gained considerable attention over the years for their predictive performance and self-adaptability in many complex problems. They are therefore perceived, by some, as better modeling tools than legacy statistical models \cite{yu_neural-network-based_2009, ahmed_empirical_2010, shen_novel_2020}. Artificial Neural Network (ANN) models are specifically regarded as some of the most accurate and universal models in time series modeling in areas including engineering, economy, and finance \cite{kohzadi_comparison_1996, khashei_artificial_2010, zhang_neural_2005}. As universal function approximators, they are a strong competitor for non-linear data modeling \cite{kim_forecasting_2018, koo_hybrid_2022, ciuca_layered_1997, schafer_recurrent_2006}. In this work, we focus our attention on a special case of ANN, the Long-Short Term Memory (LSTM) model, which has shown superiority in prediction accuracy, universality, and adaptivity in volatility modeling and many other time series forecasting tasks \cite{kim_forecasting_2018, gamboa_deep_2017, hochreiter_long_1997, maknickiene_application_2012, chen_lstm-based_2015}. However, one common problem of the ANN models, including the LSTM model, is overfitting, where the model adheres too closely to in-sample data and loses generalizability on out-of-sample predictions. 

Our primary contribution draws inspiration from Physics Informed Machine Learning (PIML), an emerging class of ML that directly embeds physical laws into the architecture or loss function of a model, thereby improving generalizability and veracity \cite{raissi_physics-informed_2019, karniadakis_physics-informed_2021, chen_accelerating_2023}. In the same way that PIML merges ML fundamentals with physical laws, this paper merges ML with stylized facts, empirical market patterns captured by the GARCH model, which serve as the “physical laws” encoded in the model. We refer to our novel hybrid model as the  GARCH-Informed Neural Network (GINN). Specifically, the GARCH model serves as a regularization mechanism incorporated in the loss function of the ANN, guarding against overfitting. Since the GINN model learns from both the factual ground truth and the knowledge acquired by the GARCH model, we expect it to grasp both general market trends and finer details.  We hypothesize that the combination of a classic statistical approach with modern machine learning will result in a highly accurate and generalizable model. 

Four models were selected as baseline models for performance comparison, namely the GARCH model, the GJR-GARCH model, the TGARCH model, and a simple, non-hybrid LSTM Neural Network model. All the models were trained and tested on 7 representative stock market indices from across the globe. The prediction accuracy is evaluated by the Coefficient of Determination ($R^2$), Mean Squared Error (MSE), and Mean Absolute Error (MAE).

\section{Methodology}
\subsection{Volatility Process Modeling}
This paper focuses on stock market volatility, which is represented by the variance in daily log return. We use the daily closing prices to calculate daily log returns. 

Denote the daily stock close price as $P_t$, where $t$ represents time. The daily stock log return is then:
\begin{equation}
  r_t = \ln{\left( \frac{ P_t }{ P_{t-1} } \right)}
\end{equation}
Here, $r_t$ is the daily log return at day $t$, $P_t$ is the daily close stock price at day $t$, and $P_{t-1}$ is the daily close stock price at the previous day $t-1$. 

The daily log return time series $r_t$ is modeled as a linear combination of the predictable average $\mu_t$ and an unpredictable error term $\varepsilon_t$. That is, 
\begin{equation}
  r_t = \mu _t + \varepsilon _t
\end{equation}
The error term $\varepsilon_t$ consists of a normally distributed random noise $e_t$ and the conditional variance $\sigma _t^2$ that is based on past information of the time series. Different variants of the GARCH-type models have different modeling of the volatility process \cite{bollerslev_generalized_1986, glosten_relation_1993, zakoian_threshold_1994, sheppard_bashtagearch_2023}. That is, 
\begin{equation}
  \varepsilon_t = \sigma_t e_t, 
  e_t \overset{\mathrm{iid}}{\sim} N(0,1)
\end{equation}

\begin{itemize}
\item {GARCH}: 

\begin{equation}
  \sigma_t^2 = \alpha_0 
  + \sum_{i=1}^{q} \alpha_i \varepsilon_{t-i}^2
  + \sum_{j=1}^{p} \beta_j \sigma_{t-j}^2
\end{equation}

\item {GJR-GARCH}: 

\begin{subequations}
\begin{equation}
  \sigma_t^2 = \alpha_0 
  + \sum_{i=1}^{q} \alpha_i \varepsilon_{t-i}^2
  + \sum_{j=1}^{o} \gamma_j \varepsilon_{t-j}^2 I
  + \sum_{k=1}^{p} \beta_k \sigma_{t-k}^2
\end{equation}
\begin{equation}
  I=
    \begin{cases}
      1, & \text{if}\ \varepsilon _{t-j} < 0 \\
      0, & \text{otherwise}
    \end{cases}
\end{equation}
\end{subequations}

\item {TGARCH}: 

\begin{subequations}
\begin{equation}
  \sigma_t^2 = \alpha_0 
  + \sum_{i=1}^{q} \alpha_i |\varepsilon_{t-i}|
  + \sum_{j=1}^{o} \gamma_j |\varepsilon_{t-j}| I
  + \sum_{k=1}^{p} \beta_k \sigma_{t-k}
\end{equation}
\begin{equation}
  I=
    \begin{cases}
      1, & \text{if}\ \varepsilon _{t-j} < 0 \\
      0, & \text{otherwise}
    \end{cases}
\end{equation}
\end{subequations}

\end{itemize}
, where $\alpha$s, $\beta$s, and $\gamma$s are the coefficients, $\varepsilon$s are the error terms, and $\sigma$s are the variances. 

For the LSTM and GINN model, we model the daily log return time series $r_t$ as normally distributed with an average value $\mu_t$ and a standard deviation $\sigma_t$. That is, 
\begin{equation}
  r_t \overset{\mathrm{iid}}{\sim} N(\mu _t, \sigma _t)
\end{equation}
The volatility could then be predicted from past volatility values: 
\begin{equation}
  \sigma_t^2 = f (\sigma_{past}^2)
\end{equation}
For a given stock log return time series $r_t$, denote the true daily volatility as $\sigma_t^2$ and the predicted daily volatility as $\hat{\sigma}_t^2$. The aim of the time series models is to forecast the variance $\hat{\sigma}_t^2$ for the day $t$ on a rolling basis based on the past stock log return values $r_t$ of a finite time window. 

\subsection{Comparison Model Selection}
To set a baseline for contemporary time-series modeling, and a meaningful performance comparison for the new GINN model, several variants of the GARCH models were selected to be the baseline models for performance evaluation. Namely, the (1,1) GARCH model, the (1,1,1) GJR-GARCH model, and the (1,1,1) TGARCH model were selected as statistical models for comparison. They were selected for their representativeness, performance, and wide recognition in the time-series field. For simplicity, the $(p,o,q)$ values of the ARCH-like models will be dropped from the name in the following sections of the paper. All GARCH-type modeling and related calculations are done in Python with the \textit{arch} library by Kevin Sheppard \cite{sheppard_bashtagearch_2023}. A naïve LSTM model was selected as an additional baseline model, representing the modern ML models in the time-series forecasting domain. The LSTM model and the LSTM component of the GINN model are deployed, trained, and tested with the \textit{PyTorch} library \cite{paszke_pytorch_2019}.

\subsection{GARCH-Type Model Formulation}
\begin{figure}[h]
  \centering
  \includegraphics[width=\linewidth]{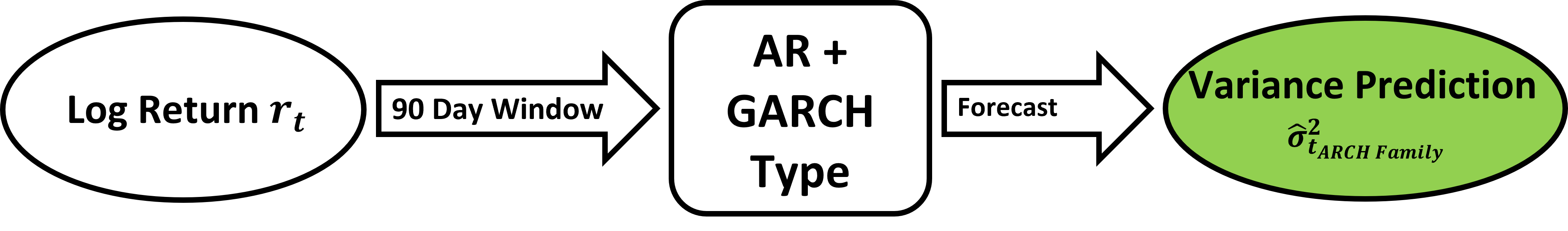}
  \caption{The variance prediction workflow diagram for GARCH-type models}
  \label{GARCHWF}
\end{figure}

The GARCH-type models construct their volatility process upon an autoregressive (AR) estimation of the average daily log return $\hat{\mu}_t$, with a daily log return time series input $r_t$. All selected GARCH-type models were configured in the same way to make rolling predictions. More specifically, the GARCH-type models would forecast the variance $\hat{\sigma}_t^2$ for the day $t$ on a rolling basis with the mean estimation from an AR model using the daily stock log return $r_t$ for the past 90 days $(r_{t-90},r_{t-89},r_{t-88},...,r_{t-1})$. Thus, we have:
\begin{equation}
    \hat{\sigma}_{t_{GARCH \enspace Type}}^2 = G (r_{t-90},r_{t-89},r_{t-88},...,r_{t-1}) 
\end{equation}
where $G$ indicates any GARCH-type model used in this work. The variance prediction workflow diagram for GARCH-type models is shown in Figure~\ref{GARCHWF}.

\subsection{LSTM Model Formulation}
\label{LSTMsection}
The procedure for the LSTM model is slightly different, as it relies on the historical ground truth variance $\sigma_t^2$ for training. The predicted daily log returns $\hat{\mu}_t$ from another model are needed to obtain the ground truth variance. This is also true for the GINN model. Here, we use the same AR model as the GARCH model to obtain predictions of the average daily log returns $\hat{\mu}_t$. The AR model forecasts the average predicted daily log returns $\hat{\mu}_t$ for the day $t$ on a rolling basis using the daily stock log returns $r_t$ for the past 90 days $(r_{t-90},r_{t-89},r_{t-88},...,r_{t-1})$. 
\begin{equation}
    \hat{\mu}_t = AR (r_{t-90},r_{t-89},r_{t-88},...,r_{t-1}) 
\end{equation}
The ground truth variance is then: 
\begin{equation}
    \sigma_t^2 = \left( r_t - \hat{\mu}_t \right) ^2
\end{equation}

\begin{figure}[h]
  \centering
  \includegraphics[width=\linewidth]{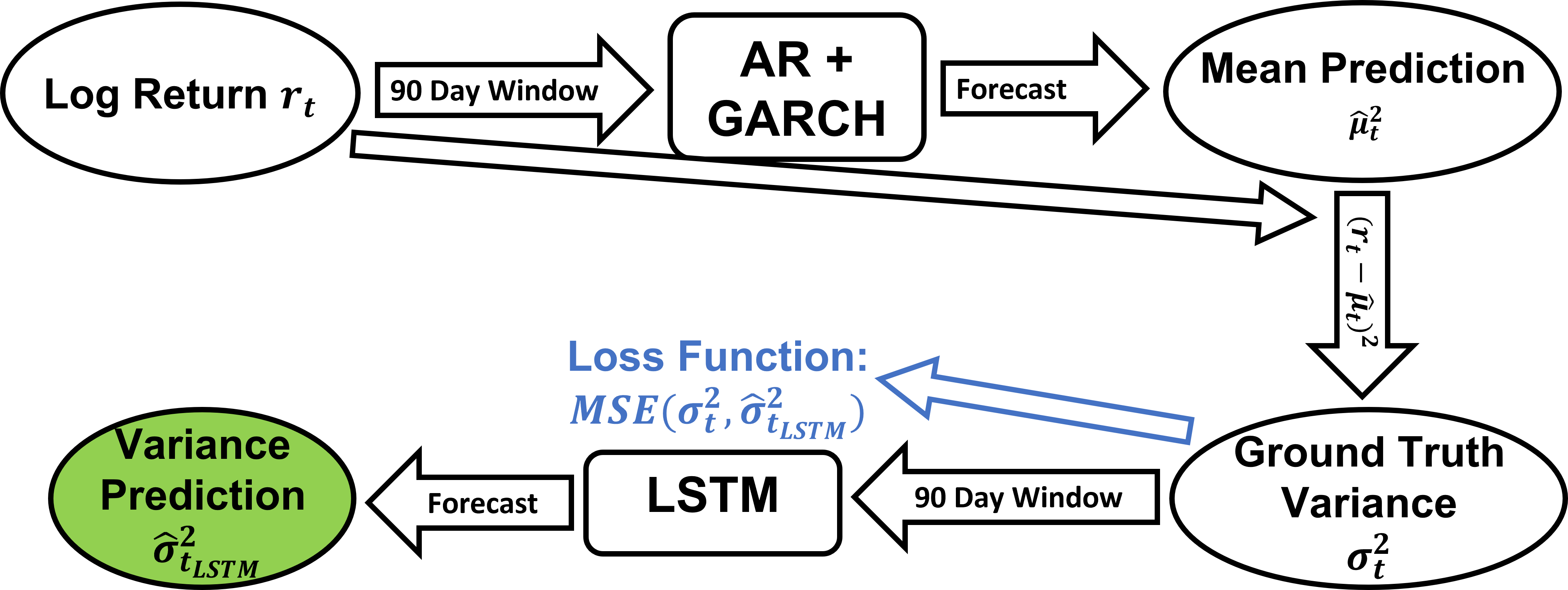}
  \caption{The variance prediction workflow diagram for the LSTM model}
  \label{LSTMWF}
\end{figure}

The LSTM model then predicts the variance $\hat{\sigma}_{t_{LSTM}}^2$ for the day $t$ on a rolling basis, using $\sigma_t^2$ for the past 90 days $(\sigma_{t-90}^2, \sigma_{t-89}^2, \sigma_{t-88}^2,...,\sigma_{t-1}^2)$. We then have: 
\begin{equation}
    \hat{\sigma}_{t_{LSTM}}^2 = LSTM (\sigma_{t-90}^2, \sigma_{t-89}^2, \sigma_{t-88}^2,...,\sigma_{t-1}^2)
\end{equation}
The variance prediction workflow chart is shown in Figure~\ref{LSTMWF}.

The LSTM model structure is shown below. The model architecture is optimized for prediction performance and represents a modern LSTM time series model with network components commonly used in similar models. The model is sufficiently large and deep, as models with more layers and parameters obtained no meaningful performance gain in our testing. The Neural Network has 3 LSTM layers with 256 layer width and Dropout layers in between, followed by two linear layers, 1 BatchNorm layer, and 1 ReLU layer as the activation function \cite{ioffe_batch_2015, agarap_deep_2018, srivastava_dropout_2014}. The model converges with the AdamW optimizer to minimize the Mean Squared Error (MSE) on predicted variance compared to the ground truth \cite{loshchilov_decoupled_2017}. 

\begin{figure}[h]
  \centering
  \includegraphics[width=\linewidth]{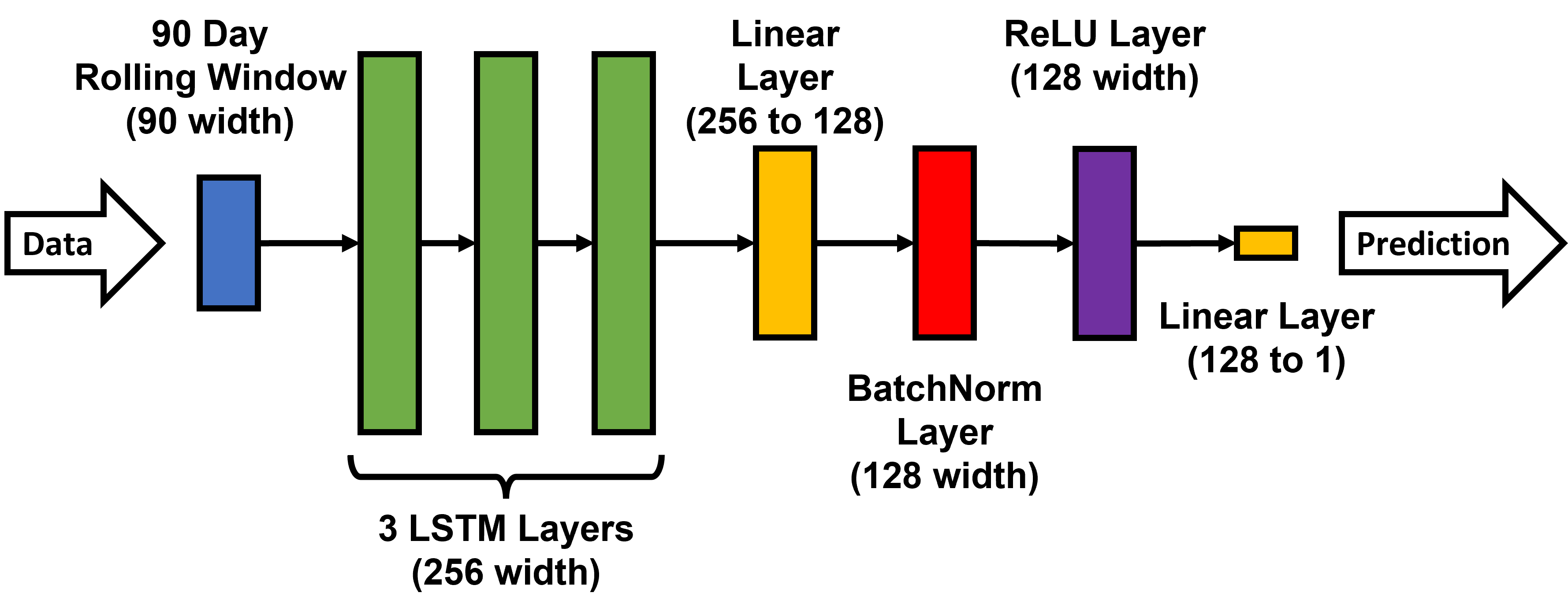}
  \caption{The model structure of the LSTM model}
\end{figure}

\subsection{GINN Model Formulation}
The proposed GINN model is a combination of a GARCH model and an LSTM model. It divides the time series forecasting task into two phases: initial prediction and calibration. The initial prediction phase uses a GARCH model, and the calibration phase is achieved with an LSTM model. Similar to the LSTM model introduced in Section~\ref{LSTMsection}, the GINN model also needs the ground truth variance $\sigma_t^2$ for training. In the initial prediction phase, an AR model and a constant mean (1, 1) GARCH model first yield the predicted average daily log return $\hat{\mu}_t$ and variance $\hat{\sigma}_t^2$ for the day $t$ in a rolling basis using the daily stock log returns $r_t$ for the past 90 days $(r_{t-90},r_{t-89},r_{t-88},...,r_{t-1})$. 
\begin{equation}
    \hat{\mu}_t = AR (r_{t-90},r_{t-89},r_{t-88},...,r_{t-1}) 
\end{equation}
\begin{equation}
    \hat{\sigma}_{t_{GARCH}}^2 = GARCH (r_{t-90},r_{t-89},r_{t-88},...,r_{t-1}) 
\end{equation}
The ground truth variance is then: 
\begin{equation}
    \sigma_t^2 = \left( r_t - \hat{\mu}_t \right) ^2
\end{equation}
In the second phase, the LSTM component of the GINN model obtains the variance prediction $\hat{\sigma}_{t_{GINN}}^2$ for the day $t$ in a rolling basis, using ground truth variance $\sigma_t^2$ for the past 90 days $(\sigma_{t-90}^2, \sigma_{t-89}^2, \sigma_{t-88}^2,...,\sigma_{t-1}^2)$. The variance prediction results from GARCH $\hat{\sigma}_{t_{GARCH}}^2$ would serve as a regularization term for the NN. 
\begin{equation}
    \hat{\sigma}_{t_{GINN}}^2 = GINN (\sigma_{t-90}^2, \sigma_{t-89}^2, \sigma_{t-88}^2,...,\sigma_{t-1}^2) 
\end{equation}

The Neural Network model structure is identical to the LSTM model, as shown in Figure 3. It contains 3 LSTM layers with 256 layer width and in-between Dropout layers, followed by two linear layers, 1 BatchNorm layer, 1 ReLU layer, and uses the AdamW optimizer \cite{ioffe_batch_2015, agarap_deep_2018, srivastava_dropout_2014, loshchilov_decoupled_2017}. 

The main difference from the naïve LSTM model is that in the GINN model, the ground truth variance $\sigma_t^2$ and the GARCH predicted variance $\hat{\sigma}_{t_{GARCH}}^2$ both act as the ground truth labels for model training. The total loss is a weighted combination of the MSE loss between $\sigma_t^2$ and $\hat{\sigma}_{t_{GINN}}^2$, and the MSE loss between $\hat{\sigma}_{t_{GARCH}}^2$ and $\hat{\sigma}_{t_{GINN}}^2$. Denote the weight as $\lambda$, we have: 
\begin{equation}
\begin{split}
    Loss = \lambda \times MSE \left( \sigma_t^2, \hat{\sigma}_{t_{GINN}}^2 \right) \\
    + ( 1 - \lambda) \times MSE \left( \hat{\sigma}_{t_{GARCH}}^2, \hat{\sigma}_{t_{GINN}}^2 \right)
\end{split}
\end{equation}

\begin{figure}[h]
  \centering
  \includegraphics[width=\linewidth]{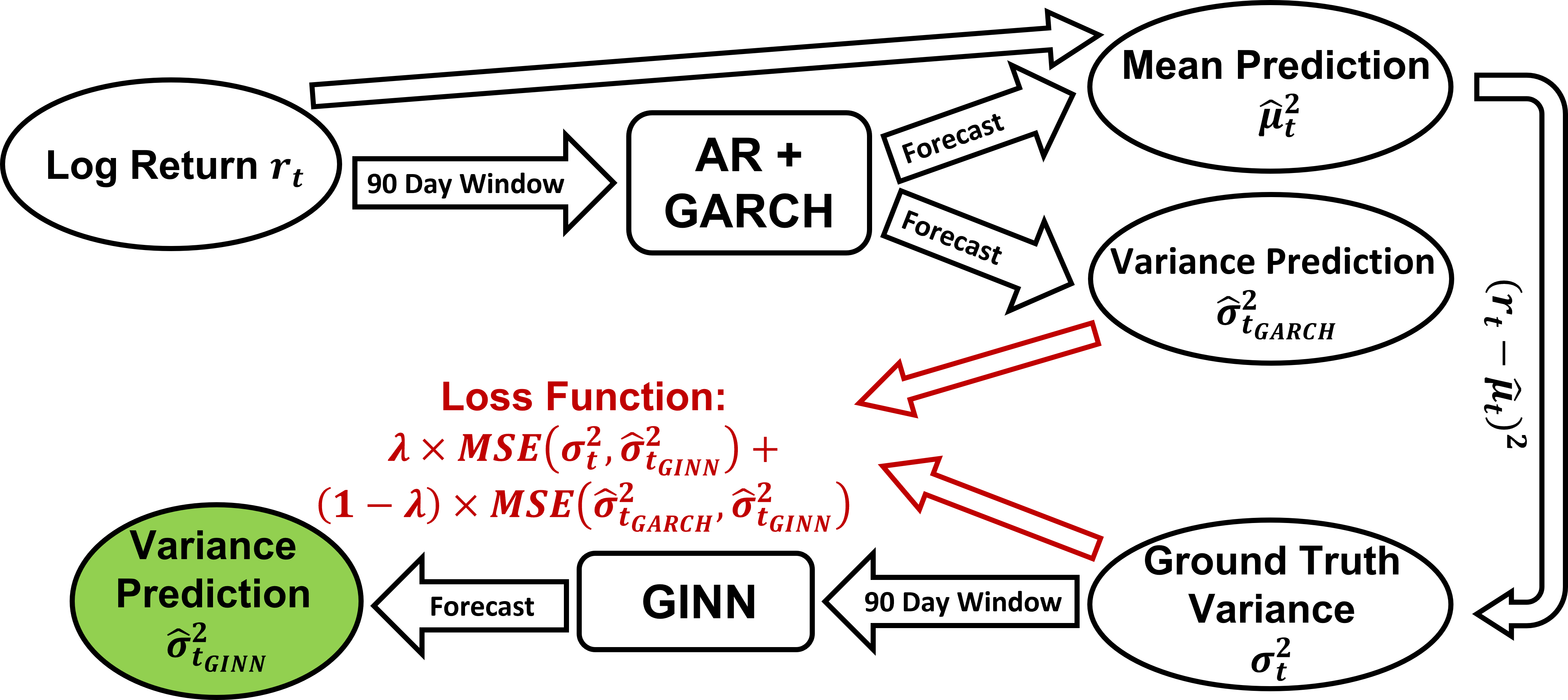}
  \caption{The variance prediction workflow diagram for the GINN model}
  \label{GINNWF}
\end{figure}

This combined loss function allows the GINN model to correct its prediction results towards a combination of the ground truth variance and the GARCH predicted results. The weight term is optimized and selected through a parametric study covered in the following subsections. The model workflow diagram is shown in Figure~\ref{GINNWF}.  

\subsection{GINN-0 Model Formulation}
Through testing, we were surprised to find that a special case of the GINN model achieved high performance. We refer to this as the GINN-0 model, indicating the GINN model with $\lambda = 0$. This means that GINN-0 only takes the volatility prediction from the GARCH model for loss calculation. That is, 
\begin{equation}
    Loss = MSE \left( \hat{\sigma}_{t_{GARCH}}^2, \hat{\sigma}_{t_{GINN}}^2 \right)
\end{equation}
Essentially, the GINN-0 model is trained to predict volatility results from the GARCH model. 

\subsection{Experiment Approach}
For training and evaluation, 7 representative stock market indices were selected from different markets across the globe: S\&P 500 Index (GSPC), Dow Jones Industrial Average (DJIA), NYSE Composite Index (NYA), Russell 2000 Index (RUT), Hang Seng Index (HSI), Nikkei 225 (NIK) and Financial Times Stock Exchange 100 Index (FTSE). 

For each aforementioned stock market index, approximately 7,500 days of daily closing values were captured from 06/01/1992 to 05/31/2022 through Yahoo Finance \cite{yahoo_yahoo_2023}. The exact number of days and the dates differ slightly from index to index, as the market opening and data availability vary across regions and indices. Each data source was then divided into a training set ($\sim 70\%$) and a testing set ($\sim 30\%$), with the date 06/01/2013 as the dividing point separating the two datasets to avoid information leakage. 

The time series models are trained on the training data and evaluated on the separate testing data, where the training data and the testing data come from the same stock market index time series. A total of 6 different models were trained and tested. Namely, the GARCH model, the GJR-GARCH model, the TGARCH model, the LSTM model, the GINN model, and the GINN-0 model were studied. All the model predictions are made on a rolling basis with a 90-day window. Each NN model was trained multiple times to reduce the uncertainty introduced by the randomness in the machine learning weight initialization process. 

To gain more insight into the relative performance of the GINN and GARCH models, we further tested them on an artificially-generated time series that explicitly satisfy the GARCH process. The GARCH model should yield equal or better performance in this simulated dataset. The volatility process of the (1, 1) GARCH model is defined as $\sigma_t^2 = \alpha_0 +  \alpha \varepsilon_{t-1}^2 +  \beta \sigma_{t-1}^2$. Different weight values of $\alpha$ and $\beta$ would result in a different GARCH time series, and could affect model performance. 

The performance of the models is evaluated by comparing the forecasted values of variance $\hat{\sigma}_t^2$ to the ground truth variance $\sigma_t^2$. Three different performance metrics are used for evaluation, including the Coefficient of Determination ($R^2$), Mean Squared Error (MSE), and Mean Absolute Error (MAE). 

\begin{enumerate}
\item \textit{Coefficient of Determination ($R^2$).} The Coefficient of Determination ($R^2$) compares the total prediction error of the model to an average value prediction, ranging from $-\infty$ to 1. Higher values indicate more accurate predictions.  A score of 1 indicates perfect predictive accuracy, 0 indicates performance identical performance to a naïve mean prediction model. 

\item \textit{Mean Squared Error (MSE).} The Mean Squared Error (MSE) is the average value of the squared prediction errors between the estimated values and the observed actual values. It measures how well the model could fit the data, with a smaller value indicating more accurate prediction results and better model performance. 

\item \textit{Mean Absolute Error (MAE).} Similar to the Mean Squared Error, the Mean Absolute Error (MAE) is the average value of the absolute prediction errors between the estimated values and the observed actual values. It measures how well the model fits the data, with a smaller value indicating more accurate predictions and better model performance. The MAE is less sensitive to extreme values than the MSE. 
\end{enumerate}

Together, the Coefficient of Determination ($R^2$), Mean Squared Error (MSE), and Mean Absolute Error (MAE) give a comprehensive measurement of the model prediction results accuracy.

\subsection{Parametric Study on the GINN model}
The performance of the GINN model is sensitive to the choice of the weight term $\lambda$. Thus, a parametric study was conducted on the weight term in order to identify the value that maximizes predictive accuracy. Different values of $\lambda$ were tested ranging from 0 to 1, with 0.01 increments between 0 and 0.2, and 0.05 increments between 0.2 and 1. 

As cross validation methods like K-folds and leave-one-out are hard to achieve with continuous time series prediction while strictly preventing information leakage, a separate and dedicated dataset was used to conduct the parametric study. This provides a safeguard against overfitting on the hyper-parameter level, and also ensures the performance obtained with the optimized weight values are generalizable to other datasets. The NASDAQ Composite (IXIC) dated from 06/01/1992 to 05/31/2022 is used as the dataset for the parametric study. The dataset is also divided into a training set ($\sim 70\%$) and a testing set ($\sim 30\%$), and fed into the prediction models in 90-day windows. The results of the out-of-sample testing set are used for performance evaluation. All three criteria, Coefficient of Determination ($R^2$), Mean Squared Error (MSE), and Mean Absolute Error (MAE), are taken into consideration. The best weight values should maximize the Coefficient of Determination, while minimizing Mean Squared Error and Mean Absolute Error. 

For each weight value, the NN model was trained multiple times to reduce the uncertainty introduced by the randomness in the machine learning weight initialization process. Both the average performance metrics results and the best performance metrics results are taken into consideration. In the case of conflicting results, the weight that yields the better average results is prioritized. Also, we look for the weight values that maximize the Coefficient of Determination ($R^2$) in the testing set when the results from the three performance metrics are inconclusive, as the Coefficient of Determination ($R^2$) is regarded as the most incisive criteria out of the three. Through testing, we determined that $\lambda=0.01$ would yield the best performance. 

It was found that both the LSTM and the GINN models reached convergence after training for 300 epochs. Since the data structure and information density is similar among the seven studied datasets, both the LSTM and the GINN models are trained for 300 epochs in all prediction tasks to ensure model convergence.

\section{Results}
\begin{figure}
  \captionof{table}{Performance Results of Models in the Out-of-Sample Testing Dataset}
  \centering
  \includegraphics[width=0.48\textwidth]{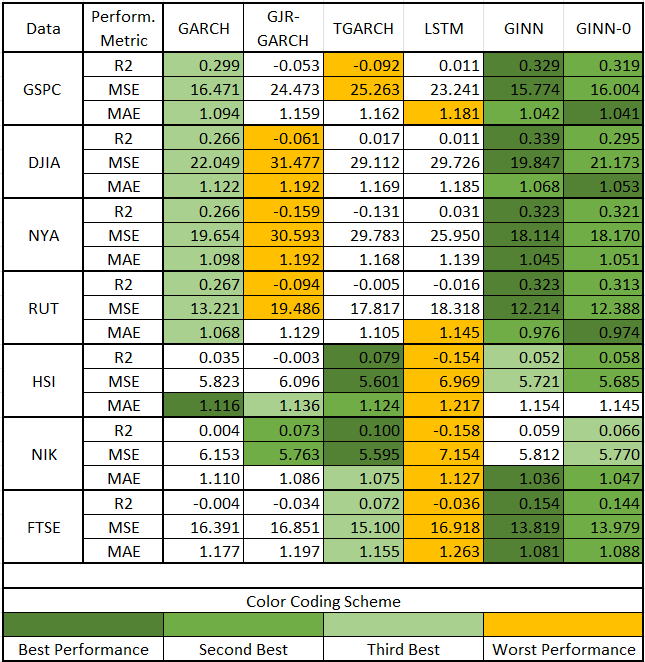}
  \label{bigoldtable}
\end{figure}

The performance results for all models in the out-of-sample testing dataset are provided in Table~\ref{bigoldtable}. For each time series data and performance metric combination, the three best performing models are highlighted with green and a darker shade stands for better performance. The worst performing model is highlighted in yellow. 

Overall, the proposed GINN model and the GINN-0 model achieved high performance in out-of-sample prediction accuracy, beating other models in most test categories. Combined, they scored 16 first places, 18 second places, and 4 third places out of a total of 21 time series data and performance metric combinations. When looking at the scores, the GINN model is $5.81\%$, $22.72\%$, $18.79\%$, and $22.05\%$ better than the GARCH model, the GJR-GARCH model, the TGARCH model, and the LSTM model, respectively, on average. The GINN-0 model is $5.43\%$, $22.38\%$, $18.45\%$, and $21.70\%$ better than the GARCH model, the GJR-GARCH model, the TGARCH model, and the LSTM model, respectively, on average. The GARCH model is the third best model in most of the out-of-sample tests, proving that it is still one of the best time series volatility prediction models with a significant lead in performance compared to the GJR-GARCH model, the TGARCH model, and the LSTM model. 

The performance scores from the GINN model and the GINN-0 model are quite close. The GINN model is only marginally better than the GINN-0 model in most of the tests, with a few incidences that the GINN-0 model surpasses the GINN model. Overall, the performance of these two models is highly comparable, and both models yield better prediction accuracy than other models, including the GARCH model. 

\begin{figure}[h]
  \centering
  \includegraphics[width=\linewidth]{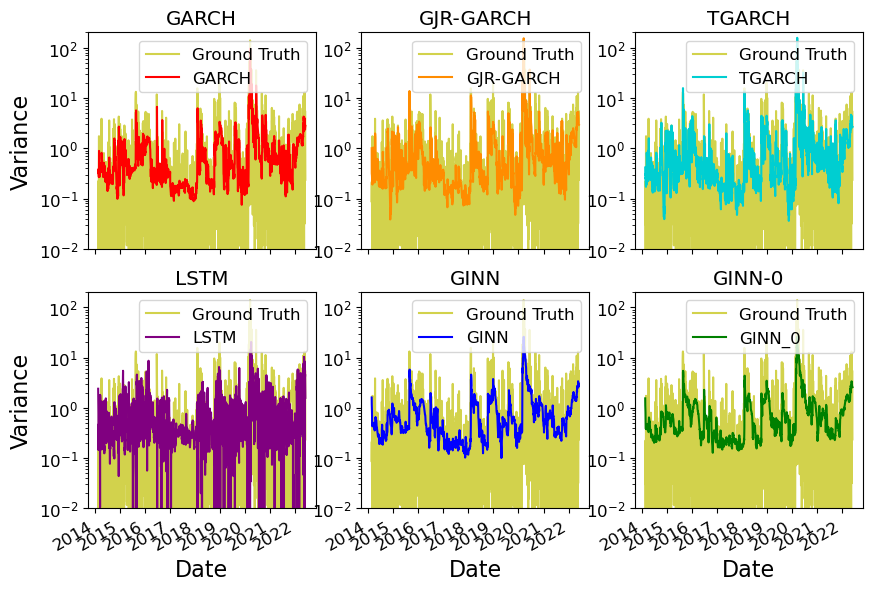}
  \caption{Daily volatility prediction results from all tested models on the out-of-sample testing set on the S\&P 500 Index (GSPC) data}
  \label{GPSC}
\end{figure}

Further, we provide examples of the out-of-sample prediction results on the S\&P 500 Index (GSPC) data from the six models against the ground truth values in Figure~\ref{GPSC}. The variance values are presented in a logarithmic scale. The horizontal axis is the date, and the vertical axis is the variance value. 

In general, the visualized results align with the numerical scores of model performance. The out-of-sample prediction results from the LSTM appear fairly random, barely capturing the larger market moves. The GJR-GARCH model and the TGARCH model successfully captured the peaks in the market (e.g., the spike in volatility in mid-2020), with seemingly more accurate magnitude prediction than the rest. However, the highly fluctuating results in the low-volatility area appear less accurate compared to the GARCH model. The GARCH results are overall smoother while preserving the market trends. The results from the GINN model and the GINN-0 fluctuate even less in comparison, to the point where it seemingly loses track of the finer details (discussed further in Section~\ref{discuss}). 

\begin{figure}[h]
  \centering
  \includegraphics[width=\linewidth]{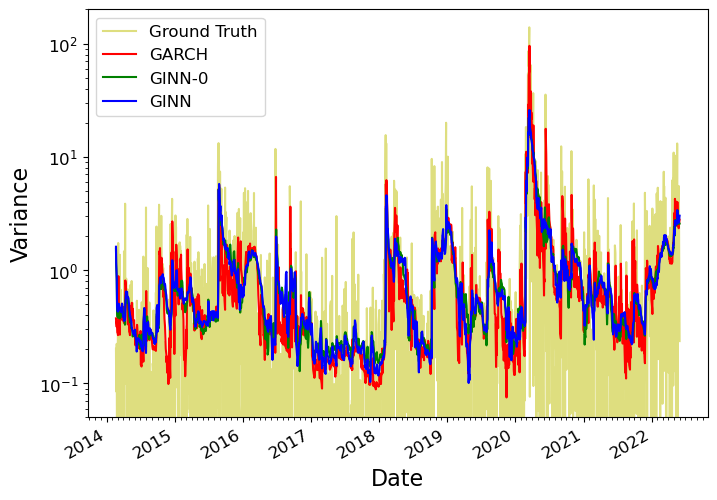}
  \caption{Daily volatility prediction results from the GARCH model, the GINN model, and the GINN-0 model on the out-of-sample testing set on the S\&P 500 Index (GSPC) data}
  \label{threebest}
\end{figure}

For easier comparison among the three best-performing models, the prediction results on the S\&P 500 Index (GSPC) data of the GARCH model, the GINN model, and the GINN-0 model are plotted in a logarithmic scale in Figure~\ref{threebest}. The horizontal axis is the date, and the vertical axis is the variance value. 

From the plot, the results from the three models look similar in shape, but with different magnitudes. The results from the GINN model look like a much smoother version of that from the GARCH model. The peak values are far less extreme, and the curve is smoother overall in day-to-day changes. The results from the GINN-0 model are highly similar to those from the GINN model. The curves from the GINN model and the GINN-0 model are hardly distinguishable, which matches the closeness in performance metrics scoring between the two. The GARCH model captures the market changes better than the GINN and GINN-0 models, as its curve matches the ground truth better with a more accurate depiction of the peak volatility locations and magnitudes especially in the highly volatile periods. 

Compared to the ground truth values, all three models did poorly. This matches the generally poor scores of Coefficient of Determination ($R^2$) with all tested models, with the GINN model being the best performer at around 0.33 on the GSPC dataset. 

As for the model performance on the simulated GARCH process data, we found that the relative performance of the GINN and GARCH models are highly related to the persistence of the time series data, and the GARCH model yields better accuracy in only about half of the simulated time series data tested. The persistence of a (1, 1) GARCH model is the property of momentum in conditional variance, and is defined as the sum of the weights $\pi = \alpha + \beta$ \cite{lamoureux_persistence_1990}. It is found that when persistence is high ($0.9 \leq \pi < 1$), the GARCH model seems to perform comparably with or outperform the GINN model, while the GINN model is more likely to outperform the GARCH model when persistence is low ($\pi < 0.9$).

\section{Discussion}
\label{discuss}
Overall, the proposed GINN hybrid model demonstrated robust stock market index prediction performance. It showed superior accuracy in the out-of-sample prediction tasks, surpassing all other tested models in most stock market indexes studied. However, upon further inspection, the performance of the GINN model may not be as dominant as it initially seems. There are many more aspects to consider in the performance of these models. 

Firstly, the performance of the GINN and GINN-0 model is qualitatively similar to the GARCH model, as expected. When the GARCH model struggles with the time series data, the GINN model and the GINN-0 model may also fail to yield accurate prediction results. This is illustrated in results from the Nikkei 225 (NIK) and the Hang Seng Index (HSI) datasets, where the TGARCH model and the GJR-GARCH model yielded low accuracy ($R^2 <= 0.1$) while still outperforming GINN, GINN-0, and GARCH. In the NIK dataset, the GINN-0 model ended up in third place, and the GINN and GARCH models placed fourth and fifth, respectively. The GARCH model's low performance led to the low accuracy of the GINN and GINN-0 models, which make direct use of GARCH. This reinforces prior work which demonstrates cases in which GARCH fails to make accurate predictions, and here those cases hinder the performance of the GINN model and the GINN-0 model \cite{franses_forecasting_1996, lee_are_1991}. Future exploration of GINN-like hybrid models should consider the use of other variants of the ARCH or GARCH-type models. 

As for the comparison between the GINN model and the GINN-0 model, the GINN model usually outperforms the GINN-0 model by a small but noticeable margin. Both models always yield better performance than the GARCH model. The GINN model's better performance could result from the fact that it learns market knowledge from both the ground truth volatility and the GARCH model. We are surprised that the GINN-0 model yields better performance than the GARCH model in some cases, even though it doesn't utilize ground truth volatility in the loss function. It is still unclear why a model trained to predict the prediction results from the GARCH model would outperform the GARCH model itself. The smoother curve and profile of the GINN-0 model compared to the GARCH model may suggest that the LSTM component of the GINN-0 model serves as additional regularization for the model, and result in a more consistent prediction. More work is needed to investigate the actual cause of GINN-0's performance gain over the GARCH model. 

Thirdly, it is noteworthy that the performance metrics used in the study may not be sufficient to picture the performance thoroughly. As mentioned previously, the prediction results from the GINN model are overly smooth compared to that from the GARCH model. It has fewer peaks and less extreme values, and lost some characteristics of the market. That somehow resulted in a better performance. It seems that the performance metrics used, the $R^2$, MSE, and MAE, tend to reward smoother prediction results. One explanation is that the less smooth models, e.g., GJR-GARCH and TGARCH, get it wrong with their variations. They over-predict and under-predict instead of taking the medium and smoothed path. Also, compared to more conservative and average-inclining values, the metrics used punish wrong and extreme prediction values more severely. Thus, when the predicted fluctuating volatility got offset by a few days from the ground truth, it could result in much lower scores even though the market trend it captures is correct. This helps explain why the seemingly more fluctuating GJR-GARCH and TGARCH models generally yield worse scores. In Figure 6, the GARCH model seems to capture the market trend better than the GINN model, with more pronounced peaks and dips, but it ended up with lower scores. This raises the concern of whether the current performance metrics are sufficient or even suitable for stock volatility prediction performance evaluation, and promotes the critical need for a further and more comprehensive study of the performance metrics in the future. 

The simulated GARCH process data results unveil more insights into the performance gain of the GINN model over the GARCH model. The GARCH model and the GINN model yield comparable performance in a time series that perfectly satisfy the GARCH process. This suggests that the superior performance of the GINN model in real-world stock market data could be a result of the GINN model capturing information in the market that doesn't satisfy the GARCH process and the GARCH model fails to catch. 

The residuals between the prediction results and the ground truth variance for the GARCH and GINN models are analyzed through their amplitude spectrum with the Fast Fourier transform (FFT) process, as shown in Figure 7. It is found that the residual of the GINN model has a higher amplitude of very low (long-term) frequencies while having a lower amplitude in the high-frequency ranges compared to that from the GARCH model. This shows that the GINN model is better at capturing daily or short-time volatility features with higher frequencies, while picking up less stagnated long-term features that could be inherently noisy and unpredictable. 

\begin{figure}[h]
  \centering
  \includegraphics[width=\linewidth]{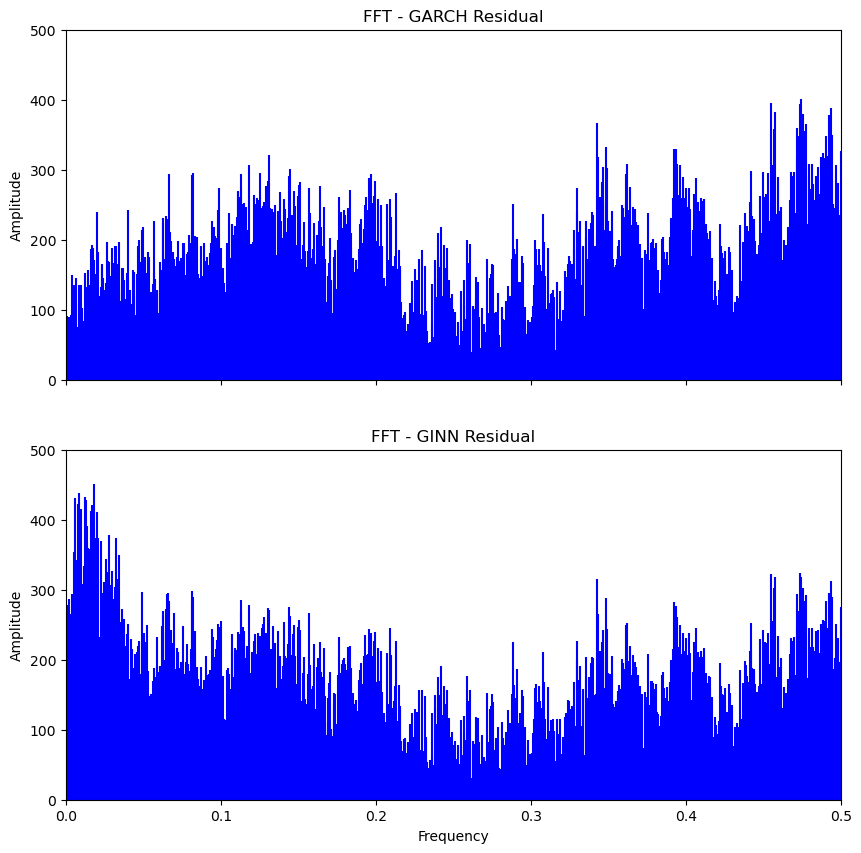}
  \caption{Amplitude Spectrum of the Residual of the GARCH and GINN models on the S\&P 500 Index (GSPC) data}
\end{figure}

Another thing that could affect the performance of the GINN model is the value of the weight $\lambda$. The current weight is selected because they result in the best average performance on a dedicated dataset to ensure the model’s generalizability. However, our testing found that some weight values could result in better accuracy in certain time series data. This leaves room for further optimization if only focusing on modeling a specific time series data. 

Lastly, all tested models did poorly in terms of prediction accuracy. One possible reason for that is the differences in the markets in the in-sample and out-of-sample time periods. The markets change over time, including the participants, regulations, and strategies in use. This may lead to the model learning relationships that are valid for older times but not for the new ages. Many unpredictable factors also contribute to market volatility, such as macroeconomic news, investor sentiment shifts, geopolitical events, and elements of market microstructure. Models based only on historical price changes may approximate volatility trends but have trouble forecasting precise volatility spikes or drops, which are often driven by these external factors. There are a multitude of sources that affect volatility in different ways, and the time series may not be autoregressive after all. The GINN model is a meaningful step toward accurate stock market modeling and prediction, but there is still much work to do before a model can accurately forecast future market trends, if that is at all possible. 

Overall, the new hybrid GINN model and the GINN-0 model showed promising performance in stock market time series prediction. They represent a new way of constructing a hybrid model. The GINN model combines the advantages of classic statistical models and modern machine learning models, learning from both the factual ground truth and the market knowledge acquired by the GARCH model. This allows the GINN model to have a better grasp of both general market trends and finer details, resulting in a highly accurate and generalizable model. The GARCH component of the GINN model also serves as an additional regularization in the loss function, working as a guard against overfitting, and improving the generalizability.

\section{Conclusion}
This paper introduced a novel hybrid model for volatility prediction in financial markets, named GARCH-Informed Neural Network (GINN). The GINN model utilized both market ground truth and volatility prediction results from the well-known GARCH model in its model training. This helps the GINN model to capture both general market trends and finer details, and results in better out-of-sample prediction accuracy in stock market volatility prediction tasks. The GINN model is noticeably better than all tested competing time series volatility models. The novel hybrid model structure introduced here provides new avenues for the construction of models for general time series modeling and forecasting.

\section*{Acknowledgment}
This material is based upon work supported by the Defense Advanced Research Projects Agency through cooperative agreement No. HR00112220029. Any opinions, findings, and conclusions or recommendations expressed in this paper are those of the authors and do not necessarily reflect the views of the sponsors. 

The authors are grateful to Nikolas Martelaro for his feedback and comments on this work.

\bibliographystyle{IEEEtranN}
\bibliography{references}
 
\end{document}